# Long-Lived Vortex Structures in Collisional Pure and Gas-Discharge Nonneutral Electron Plasmas

N. A. Kervalishvili

Andronikashvili Institute of Physics, Javakhishvili Tbilisi State University,
Tbilisi 0177, Georgia.  <n_kerv@yahoo.com>

**Abstract.** The analysis of experimental investigations of equilibrium, interaction and dynamics of vortex structures in pure electron and gas-discharge electron nonneutral plasmas during the time much more than the electron-neutral collision time has been carried out. The problem of long confinement of the column of pure electron plasma in Penning-Malmberg trap is considered. The mechanism of stability of long-lived vortex structure in gas-discharge nonneutral electron plasma is investigated. The collapse of electron sheath in gas-discharge nonneutral electron plasma in Penning cell at high pressures of neutral gas is described. The interaction between the stable vortex structure and the annular electron sheath, and the action of vortex structures on the transport of electrons along and across the magnetic field are discussed.

## I. INTRODUCTION

The previous work [1] studied the general mechanisms and the differences in the process of formation, interaction and dynamics of vortex structures in pure electron and gas-discharge electron nonneutral plasmas during a short collisionless time interval following the origination of diocotron instability. In the same work, the attention was paid to the peculiarities of experimental methods of investigation in these two plasmas. The analysis of experimental results showed that the process of formation of stable vortex structure proceeds in both plasmas practically in the same way, and the observed differences in the behavior of vortex structures connected with the different initial parameters of electron plasma. Independent of the initial number of vortex structures, at the end of the process of collisionless evolution in both plasmas only one stable vortex structure is left. However, in pure electron plasma, the vortex structure is shifted to the axis of trap, and in gas-discharge electron plasma, it remains in the electron sheath near the anode surface. The further evolution and dynamics of vortex structure takes place with the participation of electron-neutral collisions. The gas-discharge electron plasma differs from the pure electron plasma in that it exists unlimitedly long at the expense of ionization by electrons. Besides, the ionization takes place not only in plasma sheath, but also inside the vortex structure. The other characteristic feature of gas-discharge plasma is the ejection of electrons from the plasma and vortex structures along the magnetic field to the end cathodes in the form of continuous flux and periodically following pulses.

The present work deals with the analysis of the behavior of vortex structures in pure electron and gas-discharge electron nonneutral plasmas in the presence of electron-neutral collisions. The paper contains both the original oscillograms and the illustrations from other works. In section 2, the process of expansion of the column of pure electron plasma in Penning-Malmberg trap, and the problems connected with its confinement at low pressures of neutral gas are considered. In section 3, the behavior of stable vortex structure in gas-discharge electron plasma is studied at different geometries and at different pressures of neutral gas. In section 4, the mechanism of stability of long-lived vortex structure in gas-discharge electron plasma is considered. In section 5 the collapse of electron sheath in gas-discharge plasma in Penning cell at the higher pressures of neutral gas when the density of neutral plasma becomes comparable to the density of electron sheath is described. In final section 6, the interaction between the stable vortex structure and the annular electron sheath, as well as the action of vortex structures on the transport of electrons along and across the magnetic field is discussed.

## II. LONG CONFINEMENT OF PURE ELECTRON PLASMA COLUMN

A pure electron plasma is formed by injection of electrons into Penning-Malmberg trap and, therefore, its initial state can have the arbitrary given density and shape (the central column, the hollow column, several columns shifted from the axis, etc). However, independent of the initial conditions and of the consequent collisionless processes, the initial state for the time interval $\Delta t \gg \nu_0^{-1}$ ($\nu_0$ is the frequency of electron-neutral collisions) will be the axisymmetric picture with one vortex structure located on the axis of confinement device and the background of low density surrounding it. Under the action of electron-neutral collisions, the vortex structure (the electron plasma column) will be expanded. The velocity of expansion should be proportional to the pressure of neutral gas. Consequently, one could expect a strong increase of the time of plasma confinement at the transition to very low pressures. However, the experiment did not prove such suggestion. In [2], the time of confinement (the time during which the electron density at the column center was halved) of the central column of pure electron plasma in the wide range of neutral gas pressure (He, $10^{-10} < p < 10^{-3} Torr$) was measured. It turned out that for the pressure of neutral gas $p > 10^{-7} Torr$, the time of confinement is determined by the classical mobility of electrons across the magnetic field, and at lower pressures, the time of confinement does not depend any more on the pressure.

In this range of pressures the frequency of electron-neutral collisions becomes less than the frequency of electron-electron collisions. However, electron-electron collisions cannot be the reason of the observed expansion of electron column. Consequently, there is another process of radial transport of electrons that becomes dominant at low pressures. In [3] it was assumed that this process can be the asymmetry-induced transport. The asymmetry-induces transport in nonneutral plasma located in cylindrical symmetric trap is the transport of charged particles across the magnetic field occurred as a result of distortion of cylindrical symmetry caused by the imperfect construction of experimental device and by a small asymmetry of electric and magnetic fields. Though this process has been studied for a long period of time, the mechanisms of such process, as well as the agreement between the considered theories and experiment have not been fully understand yet [4].

Thus, the attempt to increase the time of confinement of electron (ion) plasma column in Penning-Malmberg trap at the expense of decreasing the neutral-gas pressure was not crowned with success. Nevertheless, the method was found allowing to counteract not only the radial expansion of electron plasma column, but to compress the electron column increasing multiply its density. This is the so-called "rotating wall" technique [5,6].

The description of stable state of nonneutral plasma in strong magnetic field, in cylindrically symmetric trap at low pressure of neutral gas is based on conservation of angular momentum. Small static asymmetries of trap construction, of electric and magnetic fields create the resistance to rotation of nonneutral plasma, and the condition of conservation of angular momentum leads to its expansion. To counteract this expansion the technique of rotating wall was developed, in which the rotating electric field is used for increasing the angular velocity of nonneutral plasma. This leads to the stabilization or to the decrease of the average radius of plasma column, i.e. to the increase of confinement time or to the radial compression of plasma. In general case, the torque from the rotating electric field will compress the plasma if the rotating electric field frequency is larger than the plasma rotation frequency [7]. Here, two regimes are possible. In [5], the frequency of applied electric field was in resonance with Trivelpiece-Gould modes and was much higher than the frequency of plasma rotation. This is the "slip" regime. In [6], the other regime was used, when the frequency of applied electric field is close to the frequency of plasma rotation. In this regime, the plasma was compressed and its density was increased until the frequency of plasma rotation approaches the fixed applied frequency. This is the "low-slip" regime. In Fig.1 taken from [6], the upper part shows the process of compression of electron column under the action of rotating wall field. The lower part of the figure shows the process of expansion of the profile of column density after the field of rotating wall is turned off. As it is seen from the figure, the rate of plasma expansion is much slower than the rate of compression.

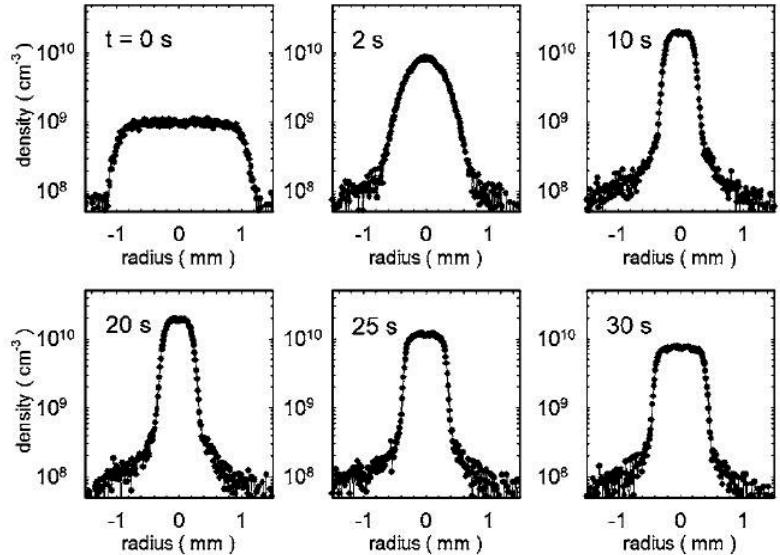

Fig.1. Evolution of the profile of electron column density
at compression (upper part) and at expansion (lower part) [6]

In this experiment, under the action of rotating wall the compressed plasma of high density was confined in the stable state for over 24 hours.

### III. STABLE VORTEX STRUCTURES IN GAS-DISCHARGE ELECTRON PLASMA

The behavior of vortex structures in gas-discharge electron plasma during the time much more than the electron-neutral collision time depends on the geometry of discharge device and on the pressure of neutral gas. In the geometry of inverted magnetron there periodically appears the diocotron instability leading to the formation of the stable vortex structure (clump). This collisionless process accompanied with the ejection of electrons along the magnetic field to the end cathodes is described in detail in [1]. Then, the formed vortex structure "decays" slowly ($\Delta t \gg \nu_0^{-1}$) [8,9]. Hence, during the most period of time we observe one quasi-stable vortex structure, the charge of which is decreased slowly. At low pressures the vortex structure has the time to disappear fully and then during some time (before appearing the next diocotron instability) the electron sheath remains undisturbed. The oscillograms of this process taken from [9] are shown in Fig.2. The upper oscillogram is the oscillations of electric field on the anode wall probe. The lower oscillogram is the full current of electrons on the end cathodes. It should be noted that in spite of electron-neutral collisions the expansion of vortex structure does not accompany its "decay". Fig.3 presents the fragments of oscillations of electric field on the anode wall probe for two time moments: when the stable vortex structure is fully formed (left oscillogram) and when it is close to full decay (right oscillogram). Here and below, the short lines on the oscillograms (to the left) indicate the initial position of the sweep trace.

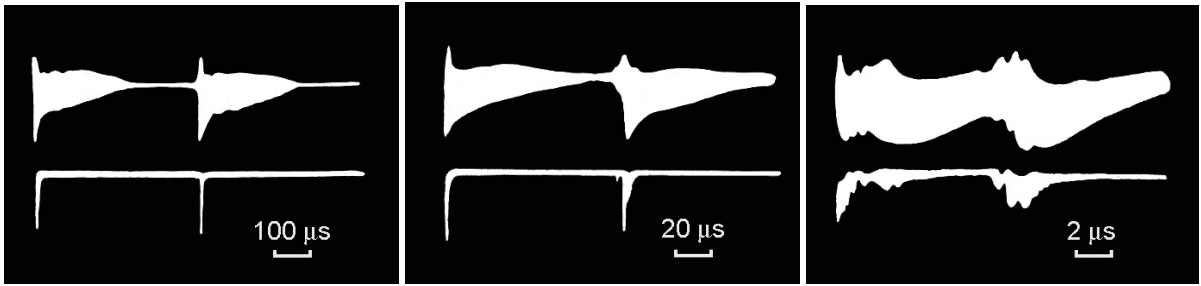

Fig.2. Diocotron instability and vortex structures in inverted magnetron [9]
$r_a = 1.0 cm$; $r_c = 3.2 cm$; $L = 7 cm$; $B = 1.8 kG$; $V = 0.9 kV$; $p = 2 \times 10^{-6}, 1 \times 10^{-5}, 1 \times 10^{-4} Torr$.

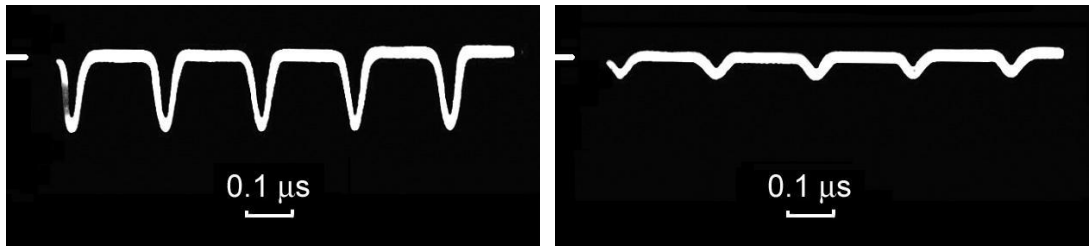

Fig.3. Decay of vortex structure in inverted magnetron
$r_a = 2.0 cm$; $r_c = 3.2 cm$; $L = 7 cm$; $B = 1.5 kG$; $V = 1.0 kV$; $p = 2 \times 10^{-5} Torr$

In magnetron geometry and in Penning cell, there exists one stable vortex structure at low pressures of neutral gas. At the pressures lower than $1 \times 10^{-5} Torr$ the drift trajectory of vortex structure periodically becomes unstable. There appear the strong radial oscillations of vortex structure that are accompanied with the ejection of electrons to the end cathodes at the moments when the vortex structure moves away from the anode surface. The period of radial oscillations of vortex structure is much more than the period of its rotation about the axis of discharge device. Therefore, during the radial oscillations the vortex structure performs a spiral motion. The orbital instability continues during the time much less than the electron-neutral collision time (5-8 radial oscillations in magnetron, and about 10 – in Penning cell). This instability is described in detail in [1,10,11]. Fig.4 taken from [10] shows the whole process of orbital instability in magnetron. The upper curve is the oscillations of electric field on the anode wall probe, and the lower one - on the cathode wall probe.

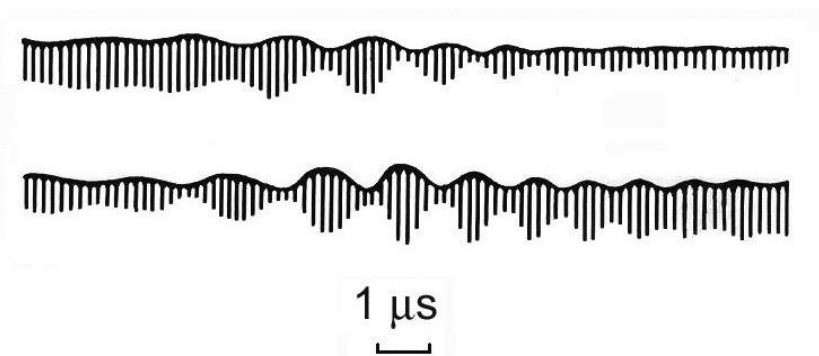

Fig.4. Orbital instability in magnetron [10]
$r_a = 3.2 cm$; $r_c = 1.0 cm$; $L = 7 cm$; $B = 1.5 kG$; $V = 3 kV$; $p = 5 \times 10^{-6} Torr$.

As a result of orbital unstability the vortex structure losses about a third of its charge and moves to the smaller orbit the radius of which is by about 5-6 mm less than the radius of vortex structure orbit before starting the orbital instability. In the interval between the orbital instabilities ($\Delta t \gg \nu_0^{-1}$) the vortex structure approaches slowly the anode increasing gradually its own charge. Fig. 5 shows the oscillograms of this process in magnetron. The upper oscillogram is the oscillations of electric field on the anode wall probe, and the lower one - the full current of electrons on the end cathodes.

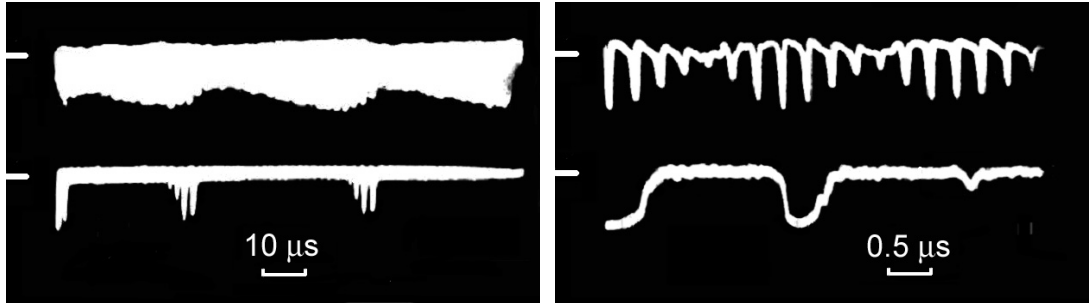

Fig.5. Periodically repetition of orbital instability in magnetron
$r_a = 3.2 cm$; $r_c = 1.0 cm$; $L = 7 cm$; $B = 1.2 kG$; $V = 1.5 kV$; $p = 6 \times 10^{-6} Torr$.

Both, the frequency of repetition of orbital instability in magnetron and the frequency of repetition of diocotron instability in inverted magnetron are proportional to the pressure of neutral gas. Both, in magnetron and in inverted magnetron, during the most period of time of periodically repeated processes the vortex structure is quasi-stable. The difference is in that in the inverted magnetron the charge of vortex structure decreases slowly, and in the magnetron – increases slowly. However, in the narrow range of neutral gas pressure, $(1-2) \times 10^{-5} Torr$, the vortex structure in the magnetron geometry remains always stable [10]. Fig.6 shows the oscillograms of oscillations of electric field on the anode wall probe (upper) and on the cathode wall probe (lower) in magnetron in this range of neutral gas pressure. As it is seen from the figure, the vortex structure does not have "tails", and its charge and orbit remain unchanged.

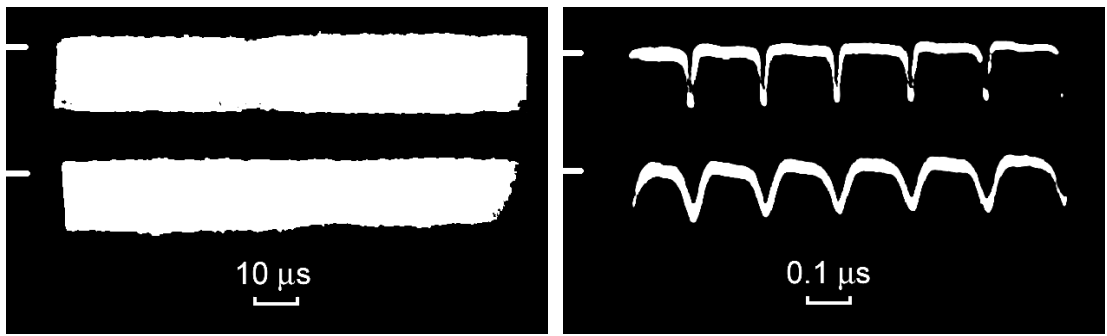

Fig.6. Stable vortex structure in magnetron
$r_a = 3.2 cm$; $r_c = 1.0 cm$; $L = 7 cm$; $B = 1.5 kG$; $V = 1.0 kV$; $p = 1 \times 10^{-5} Torr$.

Thus, in spite of electron-neutral collisions, in gas-discharge electron plasma the vortex structure can exist for a long period of time ($\Delta t \gg \nu_0^{-1}$) keeping its charge and shape.

## IV. THE MODEL OF STABLE VORTEX STRUCTURE

The stability of vortex structure at the presence of electron-neutral collisions can be connected with the simultaneous existence of two processes in the vortex structure: ionization and ejection of electrons along the magnetic field. The pulses of electron current on the end cathodes appear at the formation of vortex structures, at their approach and at radial shift of vortex structure from the anode surface [1,8-11], i.e. the moments when the local decrease of potential barrier takes place or when the vortex structure itself shifts to the region with less potential barrier. As a result, a part of electrons with the energy sufficient to overcome the decreased potential barrier goes along the magnetic field to the end cathodes.

However, besides the pulses of electron current, there is the continuous flux of electrons from the vortex structures to the end cathodes. Fig.7 taken from [12] shows the continuous flux of electrons from the vortex structures in the case of one stable vortex structure (left) and in the case of two approaching vortex structures (right). Upper oscillograms are the signals from the anode wall probe, and lower oscillograms– the current of electrons through the narrow radial slit in the end cathode. The slit is located on the same azimuth as the wall probe and the width of slit is much less than the diameter of vortex structure. As is seen from the figure, the continuous current of electrons from the vortex structure rotates together with the vortex structure around the axis of discharge device.

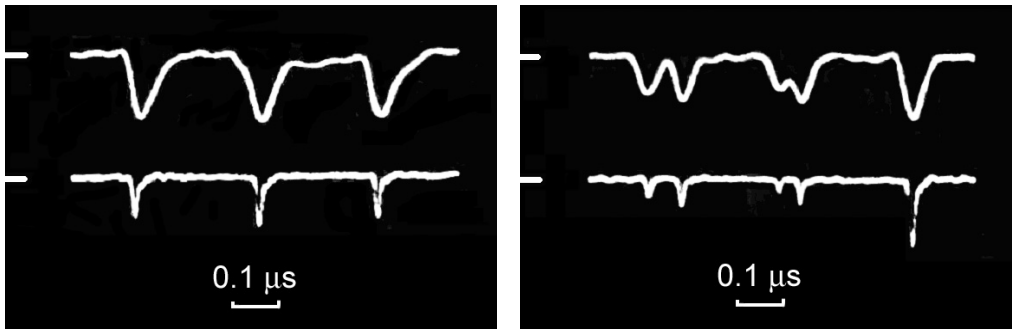

Fig.7. Continuous electron ejection from vortex structure in Penning cell [12]
$r_a = 3.2 cm$; $L = 7 cm$; $B = 1.9 kG$; $V = 1.0 kV$; $p = 1 \times 10^{-5} Torr$

The average value of total electron current on the end cathodes is rather high and make about 50% of the value of discharge current [13,14]. Therefore, this mechanism of losing the electrons is necessary to be taken into account together with the ionization and transverse diffusion at the consideration of processes taking place both, in vortex structure and in electron sheath of discharge.

In [15] the model of stable vortex structure was proposed, in which from the periphery of vortex structure on the side nearest to the cylindrical cathode, a continuous ejection of electrons takes place along the magnetic field to the end cathodes. The ejection of electrons compensates the ionization in the vortex structure and the expansion of the vortex structure due to electron-neutral collisions. Let us consider this process in more detail. The vortex structure rotates about its own axis. Therefore, the electrons of vortex structure approach periodically the anode and the cathode. The electric field near the anode is larger than near the cathode and, correspondingly, the average radial velocity of electrons near the anode is higher than near the cathode. In this case, the "longitudinal energy" acquired by vortex electrons at the expense of electron-neutral collisions near the anode, can be enough for overcoming the potential barrier near the cathode after they approach the cathode. The farther are the electrons from the vortex center, the more is the value of the "longitudinal energy" acquired by electrons near the anode and the less is the potential barrier near the cathode, and consequently, the more probable is the escape of electrons along the magnetic field. The electrons originating in the vortex structure at the expense of ionization are moved to the periphery of the structure at the expense of electron-neutral collisions and go to the end cathodes along the magnetic filed by the considered mechanism. Therefore, the transverse dimension of the structure and its charge remain unchanged. If the balance between these processes is disturbed, the charge of vortex structure will increase slowly, or on the contrary, will decrease slowly. Thus, the above-considered mechanism of the balance of processes of ionization and of escape of the electrons along the magnetic field can explain the stability of vortex structure in magnetron in the range of pressure $(1-2)\times 10^{-5} Torr$, the increase of the charge of vortex structure in magnetron at lower pressure and the decrease of the charge of vortex structure in inverted magnetron.

In the considered model the vortex structure is an isolated object not exchanging the electrons with discharge electron sheath. Nevertheless, its effect on electron sheath is quite strong. In the sheath there is the velocity shear and each electron of the sheath passes the vortex structure many times. As the vortex structure has its own electric field, the sheath electrons deviate to the anode or to the cathode while passing the vortex structure. The electrons deviated to the cathode, appear in the region of low containment potential and a part of them escape along the magnetic field to the end cathodes. Thus, alongside with the electron current from the vortex structure the current of electrons from the sheath to the end cathodes should also exist. This current will be located closer to the cylindrical cathode as compared to the current from the vortex structure. The both currents are continuous and rotate together with the vortex structure around the axis of discharge device.

Hence, in the case of one stable vortex structure the two-hump distribution of the density of electron current along the radius should be observed on the end cathode. The experimental dependences of electron current density on the discharge radius for different magnetic fields in magnetron are shown in Fig.8 in case of one stable vortex structure. The measurements were carried out by using a moving collector located behind the radial slit in the end cathode.

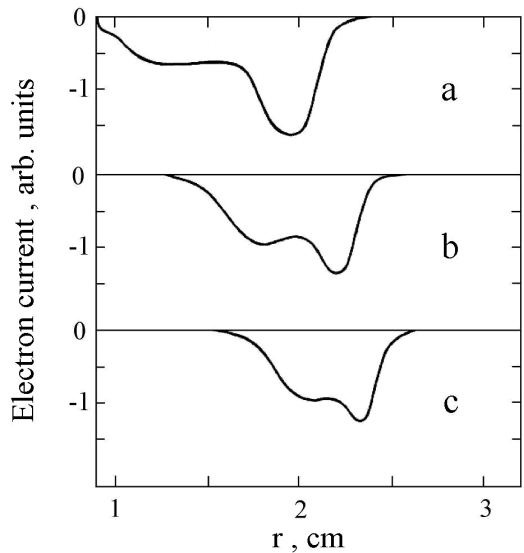

Fig.8. Electron current on the end cathode in magnetron [15]
$r_c = 0.9 cm$; $r_a = 3.2 cm$; $L = 7 cm$; $V = 4 kV$; $p = 2 \times 10^{-5} Torr$; a - $B = 1.2$, b - 1.5, c - 1.8 $kG$

As it is seen from the figure, the experimental results confirm the two-hump distribution of electron current density along the discharge radius, and thus, are in qualitative agreement with the considered model of vortex structure. It should be noted that such comparison is valid only for stable vortex structure with unchanged charge and orbit. If the charge of vortex structure is changed periodically or the vortex structure undergoes the radial oscillations, the similar measurements give the time-averaged picture "smearing" thus the real distribution of electron current along the discharge radius.

If, as a result of radial shifting the vortex structure appears in the center of Penning cell, the action of above-considered mechanism of stabilization of vortex structure will be stopped. Under the action of electron-neutral collisions the vortex structure begins to expand slowly, however, in contrast to pure electron plasma, its charge will be increased. This will lead to the formation of an annular sheath, then to the diocotron instability and further to the formation of a new off-axis vortex structure.

**V. COLLAPSE OF GAS-DISCHARGE ELECTRON SHEATH**

At higher pressure of neutral gas $\left(5\times10^{-5} < p < 1\times10^{-3} Torr\right)$, both, in magnetron and in inverted magnetron, there exist simultaneously several vortex structures, and higher is the pressure, the more is their number. This, probably, is caused by the fact that the decay rate of vortex structure increases with the pressure slower than the growth rate of electron density in the sheath. Consequently, the new vortex structure formed as a result of the next diocotron instability appears earlier than the preceding structure has time to decay. Under these conditions, the interaction of vortex structures becomes the dominant process in electron plasma. It should be noted that these simultaneously existed vortex structures are not coherent. They appear at different time, move on different circular orbits with different angular velocities [10]. The vortex structures periodically approach each other, sometimes they merge, and sometimes the new structures are formed and all these processes are accompanied with the ejection of electrons along the magnetic field to the end cathodes [8-11]. Fig.9 gives the oscillograms showing the process of approaching the vortex structures at different pressures of neutral gas. The upper oscillograms show the oscillations of electric field on the anode wall probe, and the lower – the full current of electrons on the end cathodes.

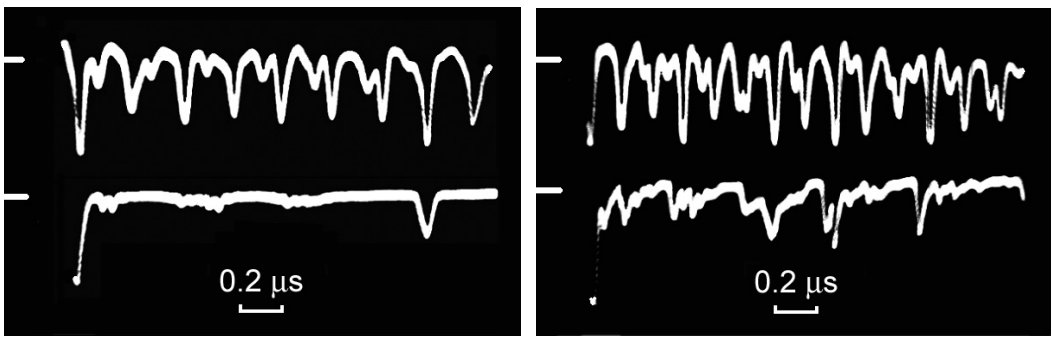

Fig.9. Approach of vortex structures in magnetron
$r_a = 3.2 cm$; $r_c = 1.0 cm$; $L = 7 cm$; $B = 1.5 kG$; $V = 1.5 kV$; $p = 1\times10^{-4}, 4\times10^{-4} Torr$

By its structure, the Penning cell is the closes analog of the Penning-Malmberg cell, as there is not a central cathode in it. However, at the pressures above $5\times10^{-5} Torr$ the average density of ions in Penning cell begins to approach the density of electrons, and in gas-discharge electron plasma in Penning cell there appear the new effects [16]. In pure electron plasma the ions are absent and such kind of problem does not arise.

In Penning cell the ions oscillate along the radius inside the hollow cylindrical anode and, at the same time, they move slowly along the axis of a cylinder towards the end cathodes. Outside the anode sheath in the central region of a Penning cell, ions are neutralized by electrons. With the increase of neutral gas pressure the density of ions increases and, as a result, becomes comparable to the density of electron sheath. This is the so-called transition mode of discharge [17-21], in which the electron sheath is "enclosed" between the anode and the neutral plasma of about the same density as the sheath. In the transition mode the discharge current shows a strong nonlinear dependence on the pressure: first, it increases rapidly, passes the maximum, then decreases, passes the minimum and then again increases sharply, and in this moment the electron sheath disappears and the discharge changes abruptly to the low-voltage glow discharge in the transverse magnetic field.

The behavior of vortex structures in transition mode was studied in [16] using a modified Penning cell, in which a flat cathode was located on the one end of cylindrical anode, as in Penning cell, and on the other end a cylindrical cathode was located, as in Penning-Malmberg cell. At such configuration, the characteristics of discharge and the behavior of vortex structures remain the same as in Penning cell with two flat cathodes. Behind the cylindrical cathode a flat collector was placed serving for measuring the current of electrons or ions ejected along the magnetic field from the electron sheath and the neutral plasma. For observation of vortex structures a diamagnetic probe was used instead of wall anode probe. This method is described in detail in [22] and consists in that only one narrow slit is cut along the whole length of the cylindrical anode, and an insulated diamagnetic probe, consisting of several turns of rf cable, is placed around the anode. When the inhomogeneity (vortex structure) passes by the slit, the image charge induced by it overcomes the slit in the anode by the current flowing around the anode circle backwards, and generating the pulse of magnetic field registered by the diamagnetic probe. The signal from the diamagnetic probe is similar to the signal from the wall probe and therefore, it is convenient to use a diamagnetic probe instead of a wall probe of external cylindrical electrode being under a high potential. Besides, a diamagnetic probe is capable to register the jumps of the magnetic field caused by the sharp changes of diamagnetic propertied of electron sheath.

The results of investigations showed that at the beginning of transition mode, the signals from collector and diamagnetic probe in the Penning cell have the same form as in magnetron. However, starting from the pressures that correspond about to a half of a discharge current maximum, the observed pattern changes significantly. In discharge, there appear the strong relaxation oscillations. Fig.10 shows the oscillograms of oscillations on the diamagnetic probe (lower), on the collector (upper left), and on the wall probe of cylindrical cathode (upper right).

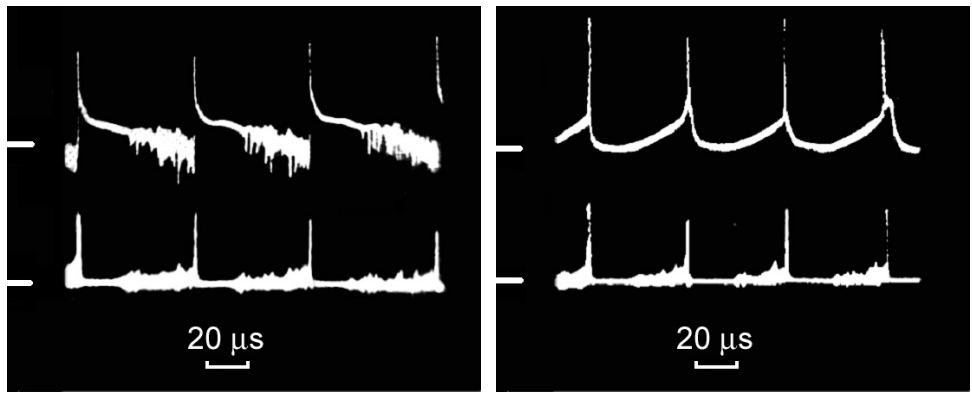

Fig.10. Collapse of electron sheath in Penning cell
$r_a = 3.2 cm$; $L = 7 cm$; $B = 1.0 kG$; $V = 1.2 kV$; $p = 1.0 \times 10^{-4} Torr$

First of all, let us pay attention to the large positive pulses on the oscillograms of the signal from the diamagnetic probe. Each pulse is caused by an abrupt increase of magnetic field due to a sharp decrease of diamagnetic properties of electron sheath. The similar pulses are observed on both sides of discharge current maximum. Fig.11 shows the oscillograms of the signal from the diamagnetic probe (lower) and of the collector current (upper) for different neutral gas pressures. The oscillograms on the left correspond to the pressures before the maximum of discharge current, and those on the right – after the maximum of discharge current.

The decrease of diamagnetic properties of electron sheath takes place as a result of ejection of a part of sheath electrons to the anode. The physical processes taking place before and after appearance of each pulse indicate this ejection. On the oscillogram of the signal from the diamagnetic probe each pulse is preceded by the oscillations connected with the motion of vortex structures around the axis of discharge device. The vortex structures are formed at the moment of appearing the diocotron instability in the discharge electron sheath as in the case of low pressures of neutral gas. The formation and the interaction of vortex structures are accompanied with the ejection of electrons along the magnetic field to the end cathodes (the upper left oscillograms in Figs 10 and 11). It should be noted that in the transition mode the formation of vortex structures is observed up to the transition to the glow discharge, while, the ejection of electrons to the end cathodes exists only up to the maximum of discharge current (Fig.11).

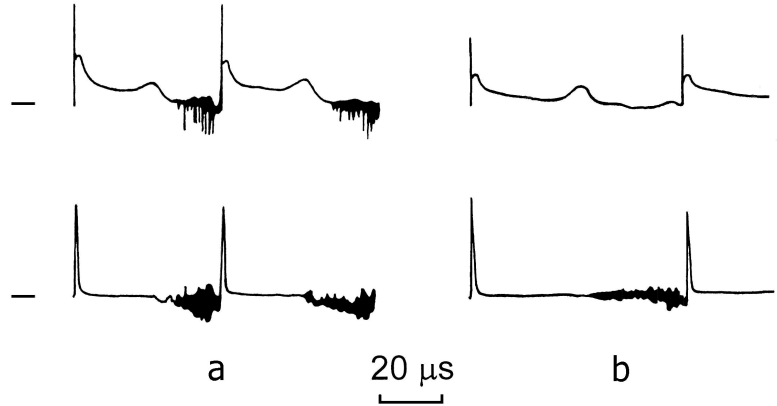

Fig.11. Collapse of electron sheath before and after the maximum of discharge current [16]
$r_a = 3.2 cm$; $L = 7 cm$; $B = 1.0 kG$; $V = 1.5 kV$; a - $p = 1.2 \times 10^{-4}$, b - $p = 1.7 \times 10^{-4} Torr$

In contrast to low pressures, when the formation of vortex structures limits the increase of electron sheath density [23,24], in the transition mode the electron sheath density continue to increase even in the presence of vortex structures, as is evidenced by a continuous increase of ion current on the cylindrical cathode (upper right oscillogram in Fig.10). The increase of electron sheath density at the fixed discharge voltage should be accompanied with its compression. This process can be continued until the sheath transforms into one-Larmor sheath, and its density reaches the Brillouin limit [25]. Such sheath is unstable. As a result of instability, the ejection of electrons takes place to the anode; this is confirmed by the fact that at the moment of jumping of magnetic field, the ejection of electrons to the collector does not take place (the left oscillograms in Fig.11). After the maximum of discharge current (the right oscillograms in Fig.11) the ejection of electrons to the collector does not take place at all. At the same time, at the moment of the jumping of magnetic field, the narrow pulses on the oscillograms of the signal from the collector and from the wall probe are observed. These pulses are caused by an abrupt increase of electric field on the cathode. (On the screened collector such pulses do not present). This confirms as well the distortion of electron sheath. The appearance of longitudinal electric field causes the sharp increase of ion current on the collector, and correspondingly, the decrease of ion current on the cylindrical cathode. Then, the electron sheath begins to recover. Its density increases as is evidenced by the increase of ion current on the cylindrical cathode (upper right oscillogram in Fig.10). Further on, there appears the diocotron instability and the whole process is repeated. Thus, we have a periodically repeated process: the compression of electron sheath and its subsequent rapid distortion, i.e. the collapse of electron sheath. The fact that with the increase of neutral gas pressure the electron sheath in Penning cell becomes thinner was observed long ago [17]. However, the interpretation of this phenomenon was different. It was assumed that the continuous (smooth or abrupt) transition takes place from the anode potential drop to that of cathode. In fact, the electron sheath in the transition mode "runs" periodically the complete cycle of its development from the formation of anode sheath to its maximum compression and subsequent destruction.

## VI. CONCLUSION

Thus, there is a certain difference in the behavior of vortex structures in pure electron and gas-discharge electron nonneutral plasmas in time interval $\Delta t \gg \nu_0^{-1}$, when the electron-neutral collisions play a significant role. In pure electron plasma the vortex structure is located on the axis of experimental device (on-axis state) and is expanded continuously until it disappears fully. For its long confinement and compression, it is necessary to use the rotating electric field - "rotating wall" technique. In gas-discharge electron plasma the vortex structure is located near the anode surface (off-axis state). This is a stable, self-organizing, and self-balancing structure retaining its charge and shape for a long period of time. In a certain sense, the vortex structure in gas-discharge electron plasma can be considered as a soliton-like structure, inside of which the formation of particles and on the periphery their removal take place.

In general, the gas-discharge electron plasma is a symbiosis of quasi-stable off-axis vortex structure and of annular electron sheath. This combination is characterized by a strong mutual effect of both components on each other and has the new surprising properties. One of them is the ejection of electrons along the magnetic field from vortex structures and neighboring regions of electron sheath to the end cathodes. The ejection is quite large and is a new unusual mechanism of losses of electrons along the magnetic field. In the vortex structure, the ejection current compensates the ionization and the expansion of vortex structure at the expense of electron-neutral collisions. In the electron sheath this current limits the electron density. In [23,24], the model of electron sheath was considered, in which the equilibrium density of electrons is determined not by a balance between the ionization and the mobility of electrons across the magnetic field, as it was assumed earlier, but by a "critical" electron density, at which there appears the diocotron instability generating the vortex structures. This model describes well the current characteristics of discharge in the crossed electric and magnetic fields both, in magnetron geometry and in the geometry of inverted magnetron [24].

The vortex structures located in electron sheath have an influence on the transport of electrons across the magnetic field too. Even in the case of one stable vortex structure, the sheath electrons pass the vortex structure multiply during the mean free time. As the vortex structure has its own electric field, the sheath electrons passing the vortex structure by, experience the radial deviations. When the radial displacement of electrons begins to exceed their Larmor radius, there appears the possibility of neoclassical transport of electrons across the magnetic field. In the process of formation of vortex structure or vortex crystals in the electron sheath (column), the turbulent dynamics appears periodically. At relatively high neutral gas pressures ($p > 10^{-4} Torr$) in the discharge electron sheath the several vortex structures exist simultaneously, and the higher is the pressure, the more is its number. At the large number of vortex structures, their motion and interaction becomes chaotic that can lead to the turbulent transport of electrons across the magnetic field.